# A spectroscopic quadruple as a possible progenitor of sub-Chandrasekhar Type Ia supernovae

Thibault Merle*[1], ORCID 0000-0001-8253-1603

Adrian S. Hamers[2], ORCID 0000-0003-1004-5635

Sophie Van Eck[3]

Alain Jorissen[4]

Mathieu Van der Swaelmen[5]

Karen Pollard[6]

Rodolfo Smiljanic[7]

Dimitri Pourbaix[8]

Tomaž Zwitter[9]

Gregor Traven[10]

Gerry Gilmore[11]

Sofia Randich[12]

Anaïs Gonneau[13]

Anna Hourihane[14]

Germano Sacco[15]

C. Clare Worley[16]

[1] Institut d'Astronomie et d'Astrophysique, Université Libre de Bruxelles, CP. 226, Boulevard du Triomphe, 1050 Brussels, Belgium

[2] Max-Planck-Institut für Astrophysik, Karl-Schwarzschild-Strasse 1, 85741 Garching, Germany

[3] Institut d'Astronomie et d'Astrophysique, Université Libre de Bruxelles, CP. 226, Boulevard du Triomphe, 1050 Brussels, Belgium

[4] Institut d'Astronomie et d'Astrophysique, Université Libre de Bruxelles, CP. 226, Boulevard du Triomphe, 1050 Brussels, Belgium

[5] INAF - Osservatorio Astrofisico di Arcetri, Largo E. Fermi, 5, 50125 Firenze, Italy

[6] School of Physical and Chemical Sciences, University of Canterbury, Private Bag 4800, Christchurch 8140, New Zealand

[7] Nicolaus Copernicus Astronomical Center, Polish Academy of Sciences, ul. Bartycka 18, 00-716, Warsaw, Poland

[8] FNRS, Institut d'Astronomie et d'Astrophysique, Université Libre de Bruxelles, CP. 226, Boulevard du Triomphe, 1050 Brussels, Belgium

[9] Faculty of Mathematics and Physics, University of Ljubljana, Jadranska 19, 1000, Ljubljana, Slovenia

[10] Faculty of Mathematics and Physics, University of Ljubljana, Jadranska 19, 1000, Ljubljana, Slovenia

[11] Institute of Astronomy, University of Cambridge, Madingley Road, Cambridge CB3 0HA, United Kingdom

[12] INAF - Osservatorio Astrofisico di Arcetri, Largo E. Fermi 5, 50125, Florence, Italy

[13] Institute of Astronomy, University of Cambridge, Madingley Road, Cambridge CB3 0HA, United Kingdom

[14] Institute of Astronomy, University of Cambridge, Madingley Road, Cambridge CB3 0HA, United Kingdom

[15] INAF - Osservatorio Astrofisico di Arcetri, Largo E. Fermi 5, 50125, Florence, Italy

[16] Institute of Astronomy, University of Cambridge, Madingley Road, Cambridge CB3 0HA, United Kingdom

* Corresponding author: tmerle@ulb.ac.be

**Binaries have received much attention as possible progenitors of Type Ia supernova (SNIa) explosions [1], but long-term gravitational effects [2, 3] in tight triple or quadruple systems could also play a key role in producing SNIa. Here we report on the properties of a spectroscopic quadruple (SB4) found within a star cluster: the 2+2 hierarchical system HD 74438 [4]. Its membership in the open cluster IC 2391 makes it the youngest (43 My) SB4 discovered so far and among the quadruple systems with the shortest outer orbital period. The eccentricity of the 6 y outer period is 0.46 and the two inner orbits, with periods of 20.5 d and 4.4 d, and eccentricities of 0.36 and 0.15, are not coplanar. Using an innovative combination of ground-based high resolution spectroscopy [5, 6, 7] and Gaia/Hipparcos astrometry [8, 9, 10, 11], we show that this system is undergoing secular interaction that likely pumped the eccentricity of one of the inner orbits higher than expected for the spectral types of its components. We compute the future evolution of HD 74438 and show that this system is an excellent candidate progenitor of sub-Chandrasekhar SNIa through white dwarf (WD) mergers. Taking into account the contribution of this specific type of SNIa better accounts for the chemical evolution of iron-peak elements in the Galaxy than considering only near Chandrasekhar-mass SNIa [12].**

The integrated spectral type of the HD 74438 system is A5; it belongs to one of the closest young clusters, IC 2391 containing 254 stars [13], located at $146^{+8}_{-7}$ pc in the Vela constellation, aged $43^{+15}_{-7}$ My [14] and of solar metallicity ([Fe/H] = –0.03 ± 0.02) [15]. It was expected that this system should at least be a triple star because it lies 0.9 mag above the main sequence (MS) of the Hertzsprung-Russell (HR) diagram of the parent cluster [16]. It was detected as a four-component spectroscopic system (SB4) only recently [4] within the ground-based, large spectroscopic Gaia-ESO Survey (GES) [5]. Follow-up observations with HRS/SALT [6] and HERCULES/UCMJO [7] (Extended Data Figure 1) allowed us to derive the orbital solution independently for the brighter (components A and B) and the fainter (components C and D) pairs (Extended Data Figure 2, Extended Data Figure 3 and Methods). The component mass ratios imply similar brightnesses in each of the inner pairs, well in line with their detection as SB2s.

We matched synthetic spectra (see Methods) to two HRS/SALT spectra where four well-separated components can be identified. One of these observed spectra is shown in Figure 1, together with the synthetic spectra of the four components and the resulting combined synthetic spectrum. The derived effective temperatures and spectral types are listed in Extended Data Figure 4.

Masses, luminosities and radii are then inferred from a stellar isochrone [17] of 43 My at solar metallicity, since the system components necessarily lie (provided no mass transfer has occurred) on the cluster main sequence (see Figure 2 and Methods). An important validation of our method is that the spectroscopic mass ratios of the outer orbit ($q_{spec} = (M_C+M_D)/(M_A+M_B) = 0.58 ± 0.11$) thus derived are in good agreement, within the uncertainties, with the dynamical ones ($q_{dyn} = 0.692 ± 0.003$, Methods and Extended Data Figure 3). Adding together the four component luminosities, a total luminosity of 15.7 ± 1.8 $L_\odot$ is obtained, which is in excellent agreement with the independent determination provided by Gaia DR2 [8, 18] ($15.3^{+0.2}_{-0.1}$ $L_\odot$, Figure 2).

We can deduce, following Equations (1), (5) and (6) of Methods, the orbital inclinations with respect to the sky plane : $i_{AB}$ = (52.5 or 127.5) ± 1.5° for the massive pair and $i_{CD}$ = (84.0 or 96.0) ± 0.9° for the low-mass pair (Extended Data Figure 3). The two SB2 orbits are thus far from being coplanar. The CD pair should moreover show eclipses that could however not be detected from the inspection of the TESS photometric data.

Thanks to archival ESO observations and our recent radial velocity follow-up, it became possible to constrain the mutual orbit of the two pairs (Extended Data Figures 5 and 6). The period and eccentricity of the wide pair are found to be 5.7 y and 0.46 (Methods and Extended Data Figure 3). The systemic velocity of the quadruple ($v_0$ = 14.5 ± 0.2 km s$^{-1}$ in Extended Data Figure 3) is in good agreement with the velocity of the parent cluster (14.98 ± 0.17 km s$^{-1}$ [19]). Moreover, thanks to the knowledge of the masses, the inclination of the wide pair on the sky is found to be (73.2 **or** 106.8)° ± 2.7°. However, the current mutual inclinations ($\Phi_{AB\,/\,AB\text{-}CD}$ and $\Phi_{CD\,/\,AB\text{-}CD}$) are not known because they require the knowledge of the longitudes of the ascending nodes $\Omega$ (Equation (12) of Methods), which can only be determined by astrometric or interferometric measurements, should they be able to turn HD 74438 in a visual system. In the Methods section, we show that Hipparcos [10, 11] and Gaia [8, 9] astrometric data succeed in doing so for the outer orbit, giving access to the corresponding longitude of the ascending node (Extended Data Figure 7). However, a similar application of the astrometric method to the inner orbits is impossible because of the closeness of the pairs, and in this case, even an interferometric imaging approach is challenging. This innovative combination of Hipparcos and Gaia astrometric data will be especially useful in the context of the forthcoming Gaia data releases.

Concerning its birth condition, HD 74438's properties may be compared to those of the much younger quadruple system discovered in the star-forming core Barnard 5 and consisting of a young protostar and three gravitationally-bound dense gas condensations [20]. The latter system has a much wider separation, but it is recognised that separations tend to decrease over time as the protostar grows in mass and dynamically interacts with the local gas reservoir [21]. These two systems thus represent different stages in the evolution of multiple-star systems within their birth environment.

HD 74438 is a rare case with a relatively short outer period of 5.7 y when compared with the other known 2+2 quadruples from the Multiple Star Catalogue [22]. Such systems are uncommon in open clusters (OCs); actually, our system has the shortest outer period known so far in an OC, as shown in the left panel of Figure 3, μ Ori [23] being the previous record holder (see also the Notes of Supplementary Information). It is currently unclear whether this results from an observational bias or from the difficulty to form them [24]. Orbital shrinking caused by gas accretion and dynamical friction mechanisms that took place during core fragmentation [21] seems to be required to account for such short outer periods.

The high level of characterisation of this quadruple system allows us to use it as a test bench to get insight into the secular evolution of multiple hierarchical systems [25]. Long-term gravitational effects in which orbit-averaged torques change the orbital angular momenta and eccentricities play a key role in triple and quadruple systems; they can indeed drive the eccentricity to high values in quadruple systems. Despite its importance, the Zeipel-Lidov-Kozai (ZLK) mechanism [2, 3] was ignored for many years, but was revived some 20 years ago for triple systems with the detection of the eccentric planet 16 Cyg B [26] or the close to perpendicular orbits in the Algol systems [27]. The double astrometric binary μ Ori has also been reported to experience ZLK cycles [23]. Within triple systems, the inner and outer orbits exchange angular momentum, which leads to mutual inclination and eccentricity oscillations on timescales much longer than the orbital periods. The inner orbit eccentricities can reach very high values, thus leading to nearly radial motion: together with orbit shrinkage, it can lead to merging of a pair [25]. In quadruple systems, orbit-averaged torques also change the orbital angular momenta and eccentricity vectors, and ZLK effects are actually even more complex than in triples [28, 29]. In systems with 2+2 architecture like HD 74438, mutual ZLK cycles can take place: each binary acts as a distant perturber on the other pair. Of particular interest is the fact that quadruple systems with mutual ZLK cycles have been proposed as one possible channel to SN Ia progenitors [30, 31], since high eccentricities are easily reached.

In HD 74438, evidence for ZLK oscillations caused by the interaction between the CD pair and the AB-CD pair is given by the fact that the eccentricity of the CD pair (0.15 in a 4.4 d orbit) is higher than what is expected for SB2 (from the SB9 catalogue [32]) and doubly eclipsing binaries [33] of similar spectral types, as clearly shown on the bottom right panel of Figure 3. Moreover the threshold period for the circularisation of binaries in clusters aged 43 My is about 7-8 d [34, their Figure 2], larger thus than the 4.4 d period of the CD pair. The CD pair should thus have been already circularised given the age of IC 2391 unless secular evolution prevented it, through the exchange of angular momentum between the outer AB-CD orbit and the CD one.

To explore the evolution of HD 74438 (up to 10 Gy in the future), we use the Multiple Stellar Evolution code (MSE) that takes into account a wide range of processes, most importantly gravitational dynamics, stellar evolution, and binary interactions such as mass transfer and common-envelope evolution [35]. We employ a Monte-Carlo approach to sample a set of realisations of HD 74438 taking into account the observational uncertainties (see Methods and Extended Data Figure 3 and Extended Data Figure 8 and 9). The first interesting result is the fact that ZLK oscillations do indeed occur in the CD pair (right panel of Figure 4), due to angular momentum transfer between the CD and AB-CD pairs, for a very wide range of mutual orbital inclinations between these pairs. Incidentally, the merging of the AB pair is even more probable (left panel of Figure 4) because the ZLK timescale of AB is shorter than the one of CD. Hence for systems like HD74438 one or more mergers occur in almost 50% of the simulated cases. We show in the Supplementary Table 4 the relative fractions of all mergers in our simulations (collisions, as well as common envelope – CE – evolution). Collisions between main-sequence stars are common, as well as CE events involving giant stars. Also possible are WD-WD mergers. The latter could lead to Type Ia supernovae [36], although in the case of HD 74438 the combined mass, according to our simulations, does not exceed the Chandrasekhar mass of 1.44 $M_\odot$ (Figure 5 and Extended Data Figure 10).

Former studies have shown that it is extremely difficult to obtain correct SNIa rates from collisions of two WDs in dense stellar environments or resulting from a merger in binary systems [30, 36 and references therein]. There were expectations that the situation would be more favourable in triples thanks to head-on collisions triggered by secular evolution involving ZLK oscillations. However, the rate of WD-WD collisions in triples has been shown to be too low to explain standard SNIa rates [36, 37]. Higher-order systems like HD 74438 provide an extra-channel towards SNIa. Nevertheless, the relative importance of the triple and quadruple channels are difficult to evaluate without detailed simulations, because the increased merger rate among quadruples [31] is possibly counter-balanced by the smaller fraction of quadruples, the exact fractions being moreover dependent upon the stellar type (Figure 39 of [39]). Even if our simulations result in sub-Chandrasekhar mass WD, it has been shown recently that between 70-85% of all SNIa could be produced by mergers leading to the explosion of a sub-Chandrasekhar WD [12, 40]. The present discovery and characterisation of a benchmark spectroscopic quadruple involving low-mass stars contributes to shed light on a channel potentially producing SNIa explosions.

## Methods

**Observations and data reduction.** The details of the observations are summarised in Supplementary Figure 3. The spectra were obtained with three different high-resolution spectrographs: (i) UVES/VLT, (ii) HERCULES/UCMJO, (iii) HRS/SALT, and with one medium-resolution spectrograph (GIRAFFE/VLT).

UVES spectra have been obtained in the context of the Gaia-ESO Survey (GES). HD 74438 has been observed 45 times within 2.5 h on the night of February 18-19, 2014, with the UVES/FLAMES multi-fiber facility using the U520 and U580 set-ups. The four components of HD 74438 were detected in the framework of a search for spectroscopic binaries [4] within the GES sample. The spectra were reduced and normalised using the GES standard reduction pipeline [41].

A long-term monitoring program of HD 74438 has been undertaken at the University of Canterbury Mt John Observatory in New Zealand with the HERCULES spectrograph [7]. The spectra were reduced and normalised with a MATLAB software pipeline developed specifically for HERCULES spectra.

Finally, monitoring programs (no. 2018-1-MLT-009 and 2020-2-MLT-003) were accepted on SALT [6] (Southern African Large Telescope) to follow with the HRS spectrograph the most interesting SB2, SB3 and the SB4 candidates uncovered within the GES [4]. The spectra were reduced with the SALT Science pipeline (http://pysalt.salt.ac.za/) [42] and the continuum normalisation was obtained by iteratively fitting polynomials and rejecting the residuals exceeding 0.25 standard deviations.

In addition, archival data from ESO with the mid-resolution spectrograph GIRAFFE/VLT allows us to add three epochs in 2004 and help to constrain the longer period but only for the AB pair because the C and D components are not resolved with GIRAFFE.

**Cross correlation function (CCF) calculations.** The CCFs were computed by cross-correlating normalised observed spectra with a unique template built as follows. The initial guess for the template was based on the spectral type attributed to the unresolved multiple system (A2). We thus adopted a Kurucz model atmosphere [43] with $T_{eff}$ = 9 000 K. Nevertheless a lower temperature ($T_{eff}$ = 7000 K) provided more contrasted CCFs. Therefore we adopted a Kurucz model atmosphere of $T_{eff}$ = 7 000 K, log $g$ = 4.0 and [Fe/H] = 0. We computed a synthetic spectrum from 3 700 to 8 900 Å to cover the three spectrographs spectral range. Strong lines were masked (Balmer lines, H&K Ca II, Ca II IR triplet, and Na I D lines). We then used Detection of Extrema (DOE) [4] to extract the positions and depths of 6 123 lines, building a comb of rectangular functions 0.01 Å-wide, and with a height corresponding to the line depth. The CCFs were then computed by cross-correlating the normalised spectra and the comb of rectangular functions.

**Radial-velocity measurements.** The identification of the number of radial velocity (RV) components, their positions and the RV uncertainties was performed with DOE [4]. First, the third derivative of the CCF was used to identify the number of components. Second, a multi-Gaussian fit was performed on the CCF. Examples of fits to multiple CCF peaks are illustrated in Extended Data Figure 1 for HRS/SALT and HERCULES/UCMJO. Three components are always visible and a fourth one can also be distinguished in most CCFs; they are labelled A, B, C, D by decreasing CCF peak intensity. While the assignment of the A and B components is straightforward, several trials and errors were necessary for a correct assignment of the C and D components because their peak intensities are similar, especially among the HRS spectra where the sparse sampling does not allow to easily follow each component over time.

**Orbital parameters of the inner pairs.** The HD 74438 system is a 2+2 hierarchical system. The time interval between UVES and HERCULES/UCMJO + HRS/SALT data is about 5 y, enough for the wider AB-CD pair to imprint a secular trend; therefore only the HERCULES/UCMJO and HRS/SALT data sets were used to derive the AB and CD short-period orbits. We also assumed that the instrumental zero-point RV offset between the two spectrographs is negligible. In Extended Data Figure 6, the final orbital solutions are compared to the HRS/SALT (middle bottom) and HERCULES/UCMJO (bottom) RVs. The symmetric ± 10 km s$^{-1}$ offsets in the GES data are well visible in Extended Data Figure 2, which displays the orbital solutions for pairs AB (left panel, $P \approx 20.6$ d) and CD (right panel, $P \approx 4.4$ d), obtained from the HRS/SALT and HERCULES/UCMJO RVs. Orbital parameters of the two short-period pairs AB and CD are provided in Extended Data Figure 3.

**Astrophysical parameters.** The astrophysical parameters were derived using two HRS/SALT spectra taken on October 14, 2018 and December 31, 2018 because they show four well-separated components at the highest resolution. Since HD 74438 belongs to a young cluster, it is assumed to be located on the main sequence and to have a solar metallicity [15] (*i.e.* log $g$ = 4.5 and [Fe/H] = 0). For the spectral fitting procedure, we built a grid of synthetic composite spectra (with each component shifted by its respective RV) covering the wavelength range [3 850, 5 500] Å computed from Kurucz model atmospheres (http://kurucz.harvard.edu/grids.html, the `ap00k2.dat` in the `GRIDP00` directory) with the radiative-transfer code Turbospectrum [44] using atomic linelist from VALD3 [45] in the range [3 850, 4 200] Å and GES linelist [46] in the range [4 200, 5 500] Å. The composite spectra in this grid combine four components, estimating radii from empirical calibration based on $T_{eff}$, log $g$ and [Fe/H] [47], with temperatures ranging from 4 000 to 10 000 K (with a step of 250 K), convolved with a Gaussian of standard deviation of 5 km s$^{-1}$. No significant broadening was detected beyond the instrumental one. The best composite synthetic spectrum was retained on the basis of the smallest standard deviation of the (observed - calculated) residuals. The adopted $T_{eff}$ of each component is the average of the values obtained from the two spectra quoted above. The $T_{eff}$ uncertainties were computed by adding quadratically the error on the mean (since the $T_{eff}$ determination was performed on two spectra, providing slightly different results for the C and D components arising from the small differences between the synthetic and observed spectra at different orbital phases) and the grid step. As an illustration of the quality of the fit, Figure 1 compares the spectrum observed on December 31, 2018 with the best combined synthetic spectrum and its individual components. The spectral types were then derived from the individual temperatures of each component. We selected three PARSEC [17] isochrones with solar metallicity (Z = 0.0147) covering the uncertainty range around the age of the parent cluster [14], namely $t = 43^{+15}_{-7}$ My. Combining these isochrones with the previously determined temperatures, the luminosities and masses of the components are derived, with uncertainties on these values deduced from the uncertainties on the component temperatures and cluster age. Radii are also derived. Finally, the spectroscopic mass ratios for each pair and for the AB-CD system, *i.e.* $M_{CD}/M_{AB}$, are derived. The astrophysical parameters of HD 74438's individual components are listed in Extended Data Figure 4. The large temperature range from the Gaia-ESO Survey (7 600 ± 750 K) well brackets the temperatures of the brightest-pair components while the temperature provided by the Gaia catalogue [8] (7 423 ± 52 K) agrees well with component B.

**Inclinations and separations.** Combining orbital and astrophysical parameters, it is possible to deduce the inclination of each SB2 pair:

$$\sin i_{\rm SB2} = \left[\frac{1}{2\pi G}\right]^{1/3} K_{\rm p} \left[\frac{P}{M_{\rm p}}\right]^{1/3} [1-e^2]^{1/2} \frac{(1+q)^{2/3}}{q} \tag{1}$$

where $K_p$, $P$, $M_p$, $e$ and $q$ are the RV amplitude of the primary, the orbital period, the mass of the primary of the pair, the eccentricity and the mass ratio of the secondary over the primary. We obtain $i_{AB}$ = (52.5 or 127.5) ± 1.5° and $i_{CD}$ = (84.0 or 96.0) ± 0.9°. Using the 3rd Kepler's law, it is possible to derive the semi-major axis of each pair in au:

$$a_{\rm SB2} = \left[\frac{M_{\rm p}+M_{\rm s}}{M_\odot}\right]^{1/3} \left[\frac{P}{1\,{\rm y}}\right]^{2/3} \tag{2}$$

where the p and s indices stand for the primary and secondary of the pair. We obtain $a_{AB}$ = 0.215 ± 0.002 and $a_{CD}$ = 0.0681 ± 0.001 au.

**Location in the Hertzsprung-Russell (HR) diagram.** The luminosity of HD 77438 (considered as a single object) is $15.3^{+0.2}_{-0.1}$ $L_\odot$ according to the Gaia DR2 catalogue [8, 18]. We now compare this luminosity to the one obtained when adding the four component luminosities. Since the temperatures of the four components have been spectroscopically determined, the cluster PARSEC isochrone at 43 My directly provides their luminosities. Summing up these luminosities leads to a total luminosity of 15.7 ± 1.8 $L_\odot$ for the combined system. It is represented by a horizontal line in Figure 2, because the temperature of the SB4 considered as a single star (for instance as given by the GES temperature of 7600 ± 750 K; blue shaded area in Figure 2) has no physical meaning. The combined luminosity is thus in excellent agreement with the one from the Gaia DR2 [18]. The above value can be compared to the luminosity also derived from the Gaia DR2 parallax [8, 13], but this time dereddening the photometry, using the usual relation:

$$L = 10^{-0.4(G+5\log\varpi+BC_G-A_G-\mathcal{M}_{\rm bol}-10)}L_\odot \tag{3}$$

where $G$, $\varpi$, $BC_G$, and $A_G$ are the Gaia $G$-band magnitude ($G$ = 7.501 ± 0.005), the parallax in mas ($\varpi$ = 6.951 ± 0.062), the bolometric correction and extinction in the $G$ band of HD 74438, respectively. The solar bolometric magnitude is taken as $\mathcal{M}_{bol}$ = 4.75. The extinction in the $G$ band depends on $G_{BP}$ - $G_{RP}$ and on the reddening for IC 2391 estimated as $E$(B−V) = 0.100 [8]. It is computed using Equation 1 and Table 1 of the Gaia Collaboration [13]: $A_G$ = 0.281$^{+0.058}_{-0.075}$ mag. The bolometric correction of HD 74438 is computed by considering the two most luminous components (A and B):

$$BC_G \approx BC_{AB}^G = BC_A^G - 2.5 \log \frac{1+L_B/L_A}{1+10^{-0.4(BC_A^G - BC_B^G)}L_B/L_A} \quad (4)$$

where $BC_{AB}^G$, $BC_A^G$ and $BC_B^G$ are the bolometric corrections in the $G$ band of pair AB, and of components A and B respectively. We found $BC_A^G$ = −0.018, $BC_B^G$ = +0.061 and $BC_{AB}^G$ = +0.010 ± 0.013. The bolometric correction model $BC_G(T_{eff})$ in the $G$ band is computed using Equation 7 and Table 4 from [18]. We finally obtain from Equation (3) an extinction-corrected luminosity of $L$ = 21.1 ± 9.5 L$_\odot$, as represented in Figure 2 by the brown dot. The uncertainty on the derived luminosity is obtained by propagating uncertainties from $G$, $\varpi$, and $BC_G$ and is admittedly large, because of the uncertainty on the bolometric correction. The cluster stars whose membership probability is higher than 90% are also represented as grey dots in Figure 2, with effective temperatures as provided by the GES iDR4 [14].

**Parameters of the outer orbit.** We retrieved observations with GIRAFFE spectrograph taken in 2004. Combined with our recent follow-up with HRS/SALT and HERCULES/UCMJO, this allows us to cover more than half of the orbital phase. We computed the centre of mass velocity of each inner pair and fit a Keplerian orbit. The fitted RV curve is given in Extended Data Figure 5. We obtained a period of about 5.7 y. The dynamical stability of hierarchical systems imposes the outer period to be at least approximately five times longer than the longest inner period [24], which is clearly the case here. The high eccentricity is not surprising for such a young system. For the derivation of the inclination of the outer pair, we used Equation (1) (but the ambiguity on the quadrant could be lifted thanks to the astrometric data discussed below) while for the semi-major axis, we directly used the third Kepler law, the values being reported in Extended Data Figure 3.

**Dynamical masses.** An independent way to derive the stellar masses (besides the location of the stars in the HR diagram) resorts to the velocity amplitudes through the set of equations:

$$\frac{M_A}{M_B} = \frac{K_B}{K_A} \qquad \frac{M_C}{M_D} = \frac{K_D}{K_C} \qquad \frac{M_A + M_B}{M_C + M_D} = \frac{K_{CD}}{K_{AB}} \quad (5)$$

The third relation leads to

$$M_C = \frac{K_{AB}}{K_{CD}} \frac{1 + K_A/K_B}{1 + K_C/K_D} M_A \quad (6)$$

which allows to derive three masses once one has been fixed. Adopting the lower possible spectroscopic mass value $M_A$ (i.e. derived from the location in the HR diagram; Extended Data Figure 4, row $M$ [M$_\odot$]$^{(a)}$), we obtained dynamical masses consistent with the spectroscopic ones for all the components. The masses derived in this way are listed in Extended Data Figure 4 in row $M$ **[M$_\odot$]**[(b)]. Because both subsystems are SB2, the dynamical mass ratios can be derived as $q = M_s/M_p = K_p/K_s$, where s and p denote the secondary and primary components, and $K$ is the RV amplitude.

**Astrometric constraints.** Astrometry from Gaia DR2 [8], Gaia eDR3 [9], Tycho-1 [10] and Tycho-2 [11] provide constraints on the orientation on the sky of the relative orbit AB-CD, as we now show. As may be seen in the $\Delta \mu_\alpha^*$ and $\Delta \mu_\delta$ rows of Supplementary Table 2, the differential proper motion of the photocentre of HD 74438 with respect to the centre of mass of the cluster varies according to both the epoch of the observations and the time span covered by the position observations used to derive the proper motion. That differential motion is compatible with zero for the Tycho-2 data since these long time-span data (about one century-long) average out the AB-CD orbital motion occurring on a much shorter time scale (5.7 y, as indicated in Extended Data Figure 3). On the contrary, the much shorter time spans of the Tycho-1, Gaia DR2 and eDR3 proper motions reveal the orbital motion on the sky. This motion is easily modelled by the Thiele-Innes constants $A, B, F, G$, which allow to express the position $(x, y)$ of a component on the sky with respect to a reference point (both will be specified below) as follows:

$y = \Delta\delta \quad\quad = A\,x' + F\,y'$

$x = \Delta\alpha \cos\delta = B\,x' + G\,y'$,

with $(x', y')$ the coordinates in the orbital plane:

$x' = r/a \cos v = \cos E - e,$

$y' = r/a \sin v = (1 - e^2)^{\frac{1}{2}} \sin E,$

and $v$ is the true anomaly, $E$ the eccentric anomaly, $e$ the orbital eccentricity, $r$ the length of the radius-vector and $a$ the semi-major axis (whose exact definition will be specified below). After derivating the above relations with respect to time, the proper motion components write

$\mu_\alpha^* \quad = \quad dx/dt \quad = n\,a''\,(B'\,\frac{dx'}{dt} + G'\,\frac{dy'}{dt}) = n\,a''(-B'\,\frac{\sin E}{1 - e\cos E} + G'\,(1 - e^2)^{\frac{1}{2}}\,\frac{\cos E}{1 - e\cos E})$

(7a)

$\mu_\delta \quad = \quad dy/dt \quad = \quad n\,a''\,(A'\,\frac{dx'}{dt} + F'\,\frac{dy'}{dt}) = n\,a''(-A'\,\frac{\sin E}{1 - e\cos E} + F'\,(1 - e^2)^{\frac{1}{2}}\,\frac{\cos E}{1 - e\cos E})$,

(7b)

with

$\dot{x}' = -\,\frac{\sin E}{1 - e\cos E}$

$\dot{y}' = (1 - e^2)^{\frac{1}{2}}\,\frac{\cos E}{1 - e\cos E}$

$A' = (\cos\omega \cos\Omega - \sin\omega \sin\Omega \cos i)$

B' = (cos ω sin Ω + sin ω cos Ω cos i)

F' = (-sin ω cos Ω - cos ω sin Ω cos i)

G' = (-sin ω sin Ω + cos ω cos Ω cos i)

being the Thiele-Innes constants divided by $n\,a"$, where $n = 2\pi / P$ and $a"$ is the orbital semi-major axis expressed in arcsec, i.e., $a" = a_{au}\,\varpi$ where $\varpi$ is the parallax in arcsec. The angles ω and Ω are the argument of periastron and longitude of the ascending node, respectively. In the following, since we are dealing with astrometric and spectroscopic orbits, these angles refer to the absolute orbit of the considered component (or photocentre) around the centre of mass of the considered system.

To go further, it is thus necessary to specify that the orbit probed by the observed differential proper motion is that of the photocentre of the AB-CD pair around the centre of mass of the system. The semi-major axis of the corresponding photocentric orbit ($a_{phot}^{AB-CD}$) is related to that of the relative AB-CD orbit ($a_{rel}$) by the relation [48]:

$$a_{phot}^{AB-CD} = a_{rel}\,(\kappa - \beta), \qquad (8)$$

with $\kappa = \dfrac{M_{CD}}{M_{AB} + M_{CD}}$

and $\beta = \dfrac{L_{CD}}{L_{AB} + L_{CD}} = \dfrac{r}{1+r}$ where $r = \dfrac{L_{CD}}{L_{AB}} = 10^{-0.4\,\Delta m}$ and $\Delta m = m_{CD} - m_{AB} \geq 0$. In the specific case of HD 74438, $a_{rel} = a_{AB} + a_{CD} = (M_{AB} + M_{CD})^{1/3}\,P^{2/3} = 5.54$ au, $\kappa = 0.41$, and $\beta = 0.0872$. Equation (8) may be converted successively in a relation for the apparent semi-major axis on the sky ($a"_{phot}^{AB-CD} = a_{phot}^{AB-CD}\,\varpi$) and in a relation for the relative orbit proper motion ($\mu_{rel} = v"_{rel} = 2\pi\,a"_{rel} / P = n\,a"_{rel}$). For HD 74438, $a"_{rel} = 36.37$ mas and $\mu_{rel} = 40.37$ mas y$^{-1}$. Thus

$\mu_{phot}^{AB-CD} = \mu_{rel}\,(\kappa - \beta) = 13.03$ mas y$^{-1}$.

From the above considerations, it results that the proper-motion components observed by the astrometric satellites (and listed in Supplementary Table 2) correspond to those of the photocentre of the AB-CD pair around the centre of mass of the system, i.e., $\mu_{phot}^{AB-CD}$, so that the value of $a"$ entering Equations (7a-b) is actually $a"_{phot}^{AB-CD}$. Therefore, one may define $\dot{X} \equiv \mu_\alpha^* / \mu_{phot}^{AB-CD}$ and $\dot{Y} \equiv \mu_\delta / \mu_{phot}^{AB-CD}$ [with three possible pairs for ($\mu_\alpha^*$, $\mu_\delta$); namely from Tycho-1, Gaia DR2, and Gaia eDR3] such that Equations (7a-b) may be rewritten, after some algebra:

$$\dot{X} = B'\,\dot{x}' + G'\,\dot{y}' = d\xi/dt\,\sin\Omega + d\eta/dt\,\cos i\,\cos\Omega \qquad (9a)$$

$$\dot{Y} = A'\,\dot{x}' + F'\,\dot{y}' = d\eta/dt\,\cos\Omega - d\eta/dt\,\cos i\,\sin\Omega, \qquad (9b)$$

with

$d\xi/dt = \cos\omega\,\dot{x}' - \sin\omega\,\dot{y}'$

dη/dt = sin ω  ẋ' + cos ω  ẏ' .

Since these various quantities are related through a rotation, it is easy to verify that

Ẋ² +  Ẏ² = (dξ/dt)²  + (dη/dt)² cos² i                                                              (10)

and

(ẋ')² + (ẏ')²  = (dξ/dt)²  + (dη/dt)² .                                                              (11)

The quantities Ẋ and Ẏ are proper-motion observables, obtained at some given epoch, whereas dξ/dt and dη/dt, as well as ẋ' and ẏ', depend on time through the orbital motion. Equations (9a-b) then allow to derive Ω without ambiguity since they provide both sin Ω and cos Ω. Since the Gaia DR2 proper motion is the one obtained on the shortest time span among all those listed in Supplementary Table 3, it most accurately represents the orbital motion and will be used in the astrometric analysis. The orbital ephemeris obtained with the orbital elements listed in Extended Data Figure 3 (especially ω = 11° and i = 73.2° or its complementary 106.8° - we note that this spectroscopic ephemeris is obviously the same for both possible values of i) predicts that the observed value Ẋ² +  Ẏ² = 0.694 occurs at temporal phase 0.022 and epoch JD 2457137 or April 24, 2015 (2015.3), in very good agreement with the mean epoch of Gaia DR2 (2015.5). At that time, the ephemeris predicts dξ/dt = -0.726 and (dη/dt) cos i = 0.415 or -0.415 (for i = 73.2° or 106.8°, respectively). From these values and the above-mentioned Ẋ, Ẏ, Equations (9a-b) then yield Ω = 333° or 274°, respectively, for the longitude of the ascending node of the orbit of the AB-CD photocentre (close to that of AB) around the AB-CD centre of mass.

The various quantities discussed above are depicted in Extended Data Figure 7, which allows to lift the ambiguity on the inclination of the AB-CD outer orbit by comparing the evolution of the proper motion from Gaia DR2 to eDR3, which has slightly turned in the anticlockwise direction, as expected for the i = 73.2° orbit, but not for the i = 106.8° one. We note that the predicted sky-projected orbital motion at the Gaia eDR3 mean epoch does not match well the observed Gaia eDR3 proper motion (dashed arrows on Extended Data Figure 7). This is because the Gaia eDR3 proper motion is an average over a large orbital span. A more rigorous approach, taking into account the exact time sampling of the orbital velocity by the Gaia scanning law and taking their average, is possible but not necessary, as we now discuss.

The possible operation of the ZLK phenomenon depends on the mutual orbital inclinations Φ (of the long orbit with respect to any of the short orbits). The angle Φ is given by the relation cos Φ = cos $i_1$ cos $i_2$ + sin $i_1$ sin $i_2$ cos ($Ω_1$ - $Ω_2$)                                        (12) where $i_1$ , $i_2$ are the inclinations of the long- and short-period orbits on the sky, and $Ω_1$ , $Ω_2$ their respective longitudes of the ascending nodes. A definite evaluation of Φ is not possible, however, in the absence of the knowledge of the longitude of the ascending node of the short orbits. Nevertheless, Equation (12) may be used to set bounds on Φ. Since -1 ≤ cos ($Ω_1$ - $Ω_2$) ≤ 1, the following inequality ensues:

cos ( $i_1$ +  $i_2$ )  ≤ cos Φ ≤ cos ( $i_1$ -  $i_2$ ).                                                    (13)

The limits on the mutual inclination derived from Equation (13) are listed in Supplementary Table 4. As shown by Figure 4, the condition 39.2° ≤ Φ ≤ 140.8° required for the operation of the ZLK process in triple systems [25] does not need to be fulfilled any longer in 2+2 quadruple systems, especially in the case of the CD pair interacting with the distant AB pair (right panel of Figure 4). Thus the ZLK process is expected to occur for all Φ values of the CD pair listed in Supplementary Table 3. Indeed, given the large eccentricity ($e$ = 0.15) for this short-period system ($P$ = 4.4 d; compare its location in the e - P diagram of similar systems with GV primaries in the bottom-right panel of Figure 3), it seems likely that the ZLK phenomenon is at work through the gravitational interaction of the CD pair with the outer AB pair.

**Birth scenario discussion.** Multiple stars are thought to be produced according to the following scenarios: (i) cluster fragmentation [49], *i.e.* core fragmentation for the outer (large period) subsystem and disk fragmentation for the inner ones (namely the 2 shorter period pairs) or (ii) dynamical and resonant capture. Several observables of HD 74438 contradict the capture scenarios. Some simulations of dynamical evolution within young clusters in the Solar Neighbourhood have initial conditions ($N$ = 182, simulation time = 50 My [50]) that match well with our case (IC 2391, with 254 objects [13], is 43 My old [14]). These simulations, including primordial single, binary, triple but not quadruple stars, under-produce the latter compared to the observed fractions by a factor of three, suggesting that a primordial population of quadruples must be present. Moreover, some magneto-hydrodynamics simulations coupled with *N*-body, stellar evolution and binary interaction codes [51] show that the binary fraction drops far below the observed statistics for low-mass stars, presumably because primordial binary formation was not included. These results tend to support a formation of lower-mass multiple systems like HD 74438 predominantly by primordial core and disk fragmentations rather than by capture within the cluster.

**Future evolution of HD 74438.** The MSE code models the evolution of multiple-star systems of arbitrary configurations [35], taking into account gravitational dynamics, stellar evolution, and binary interactions such as mass transfer and CE evolution. Also included are simple recipes for 'triple' interactions such as triple CE evolution, when a tight binary star enters the envelope of a giant star. We employ Monte Carlo methods to sample a set of realisations of the observed system taking into account the observational uncertainties. As it is not possible to disambiguate between inclinations for the two short-period orbits, we chose to only use the ones below 90°. Specifically, for each system we sample all parameters (the four masses and five orbital elements for each of the three orbits, i.e., 19 parameters in total) from Gaussian distributions centred at the observed values, and with standard deviations given by the observed error bars. An exception to the latter applies to the longitudes of the ascending nodes, $\Omega_i$, which are not all observationally constrained and which were sampled from flat distributions. With this Monte Carlo approach, we sample $N_{MC}$ = 10$^4$ systems and evolve them with MSE until a system age of 10 Gy was reached, or until the maximum allowed CPU wall time of 20 hr was reached, which occurred for about 67% of the systems. Given the highly compact nature of HD74438, the system is computationally prohibitively expensive to evolve (even using the secular approach in MSE), since the secular time-scales are short compared to 10 Gy. Most first mergers occur at an early age of ~10-100 My, significantly younger than the attained age of most of the systems for which the CPU wall time was exceeded (peaking near 10$^9$ y, see Extended Data Figure 8). This justifies the imposed maximum wall time. A summary of our results is given in Supplementary Table 4, which shows the fractions per occurring physical processes, number of remnants and outcomes of our simulations. Each simulated system can experience one or more events: one or more mergers (including both CE evolution and direct collisions), CE evolution, direct collisions, dynamical instability, and RLOF of a tertiary star onto an inner binary. In the latter case, mass transfer can proceed stably (not likely), or unstably leading to triple CE evolution. About half of realisations do not lead to interactions such as mergers, whereas the other half involves interactions such as stellar mergers, and triple CE evolution. The high fraction of triple Roche Lobe Overflow (RLOF) and triple CE (close to 40%) is much higher compared to the expectation for the entire population of isolated triple systems [52]. We classify the final outcome of the system in terms of the number of remaining objects (which can be stars or compact objects), as well as in terms of their hierarchical configuration. The non-interacting systems correspond to stable quadruple systems (fraction ~ 0.5), which are the only final configuration with four remnants (e.g., the 'Triple+Single' outcome does not occur). When three remnants remain (fraction ~ 0.1), they can be either in a stable triple (likely), or a binary with an unbound single (not likely). If there are two remnants (fraction ~ 0.1), the fractions of them being bound and unbound to each other are ~0.03 and ~0.06, respectively. A single remnant occurs for a fraction of ~ 0.2 of realisations, whereas there are no cases with no remaining remnants. In almost all instances when the tertiary fills its Roche lobe around an inner binary, the subsequent evolution is unstable and leads to triple CE evolution. In addition, the simulation show that there is a clear preference for interacting systems (in particular, mergers and triple CE systems) to have $\Phi_{AB / AB-CD}$ close to 90°, whereas the non-interacting systems favor initial $\Phi_{AB / AB-CD}$ further from 90°, with peaks near the smallest $\Phi_{AB / AB-CD}$ around 20°, and the largest around 120°, as shown in Extended Data Figure 9. This can be easily understood by noting that highly mutually inclined systems will lead to high eccentricities due to strong secular evolution [28]. The overwhelming majority (~ 98%) of first collisions indeed occur in orbit AB. On the other hand, $\Phi_{CD / AB-CD}$ shows less distinct features among the different outcomes, with the exception that non-interacting systems favour smaller $\Phi_{CD / AB-CD}$ compared to interacting ones (i.e., further separated from 90°). Extended Data Figure 10 illustrates one possible future evolution of HD 74438 towards a WD with a sub-Chandrasekhar mass: the AB pair rapidly collides after 170 My because secular evolution drives eccentricity close to unity. The merger remnant, a 3.3 $M_\odot$ MS star, subsequently evolves and fills its Roche lobe around the companion binary at 540 My. The outcome of the triple CE is an unstable system in which a collision quickly occurs between the two components of the inner binary. A WD+MS binary remains. As the MS companion evolves, it fills its Roche lobe at around 2 Gy, ultimately causing a merger. The final remnant is a single 1.3 $M_\odot$ WD.

**Data availability.** Source data are provided with this paper: the measured RVs from which all the orbital solutions are computed is available in a separate csv file (Supplementary Table 5). In addition, raw spectroscopic data are available in the associated observatory archive:

- http://archive.eso.org/scienceportal for ESO/GES and GIRAFFE spectra
- https://ssda.saao.ac.za/ for SALT/HRS spectra. The spectra from the monitoring program 2018-1-MLT-009 are publicly available except for the 2020-2-MLT-003 program (3 spectra, available from mid-2024 on).
- The HERCULES/UCMJO raw spectra will be made publicly available on the Vizier/CDS repository.

**Code availability.** The Multiple Stellar Evolution (MSE) code used to perform the simulation of HD 74438 system is available upon request to A.S.H.

## Acknowledgments

T.M. and S.V.E. are supported by a grant from the Fondation ULB. A.S.H. thanks the Max Planck Society for support through a Max Planck Research Group. T.M. acknowledges M.G. Perderson and T. Van Reeth for investigating the TESS data. T.M. acknowledges A. Tokovinin for explanations about the Multiple Star Catalogue. T.Z. and G.T. acknowledge financial support from the Slovenian Research Agency (research core funding No. P1-0188) and from the European Space Agency (Prodex Experiment Arrangement No. C4000127986). G.T. acknowledges support by the Swedish strategic research programme eSSENCE, the project grant "The New Milky Way" from the Knut and Alice Wallenberg foundation and the grant 2016-03412 from the Swedish Research Council. Based on data products from observations made with ESO Telescopes at the La Silla Paranal Observatory under programme ID 188.B-3002. These data products have been processed by the Cambridge Astronomy Survey Unit (CASU) at the Institute of Astronomy, University of Cambridge, and by the FLAMES/UVES reduction team at INAF/Osservatorio Astrofisico di Arcetri. These data have been obtained from the Gaia-ESO Survey Data Archive, prepared and hosted by the Wide Field Astronomy Unit, Institute for Astronomy, University of Edinburgh, which is funded by the UK Science and Technology Facilities Council.  This work was partly supported by the European Union FP7 programme through ERC grant number 320360 and by the Leverhulme Trust through grant RPG-2012-541. We acknowledge the support from INAF and Ministero dell' Istruzione, dell' Universit\`a' e della Ricerca (MIUR) in the form of the grant "Premiale VLT 2012". The results presented here benefit from discussions held during the Gaia-ESO workshops and conferences supported by the ESF (European Science Foundation) through the GREAT Research Network Programme. Some of the observations reported in this paper were obtained with the Southern African Large Telescope (SALT) under programs 2018-1-MLT-009 (PI: R. Smiljanic) and 2020-2-MLT-03 (PI: R. Manick). Polish participation in SALT is funded by grant No. MNiSW DIR/WK/2016/07. This work has made use of data from the European Space Agency (ESA) mission Gaia (https://www.cosmos.esa.int/gaia), processed by the Gaia Data Processing and Analysis Consortium (DPAC, https://www.cosmos.esa.int/web/gaia/dpac/consortium). Funding for the DPAC has been provided by national institutions, in particular the institutions participating in the Gaia Multilateral Agreement. This work has made use of the VALD database, operated at Uppsala University, the Institute of Astronomy RAS in Moscow, and the University of Vienna. This research has made use of the SIMBAD database, operated at CDS, Strasbourg, France. This research has made use of the VizieR catalogue access tool, CDS, Strasbourg, France (DOI : 10.26093/cds/vizier). The original description of the VizieR service was published in A&AS 143, 23. This research has made use of Python3 and IPython, and modules NumPy, SciPy and Pandas. All the graphics were generated with Matplotlib except Extended Data Figure 5.

## Author Contribution Statement

T.M. initiated the project, performed the follow-up with HRS/SALT, the analysis, the interpretation of the data and designed most figures. A.S.H. performed the simulations of the evolution of the quadruple and designed some of the figures. S.V.E. and A.J. contributed to the analysis and the interpretation of the results. T.M., S.V.E., A.J. and A.S.H. wrote the manuscript, with input from all authors. M.v.d.S. and D.P. contributed to the analysis of the data. K.P. performed the acquisition and reduction of HERCULES/UCMJO spectra. R.S. contributed to the follow-up with HRS/SALT and to the interpretation of the data. T.Z. and G.T. contributed to the analysis and interpretation of the data as well as to the writing. G.G. and S.R. are the P.I. of the Gaia-ESO survey in which the quadruple was discovered. A.G., A.H., G.S. and C.C.W. contributed to the acquisition and reduction of the GES data. All authors provided critical feedback.

## Competing interest Statement

The authors declare no competing interests.

## Figures

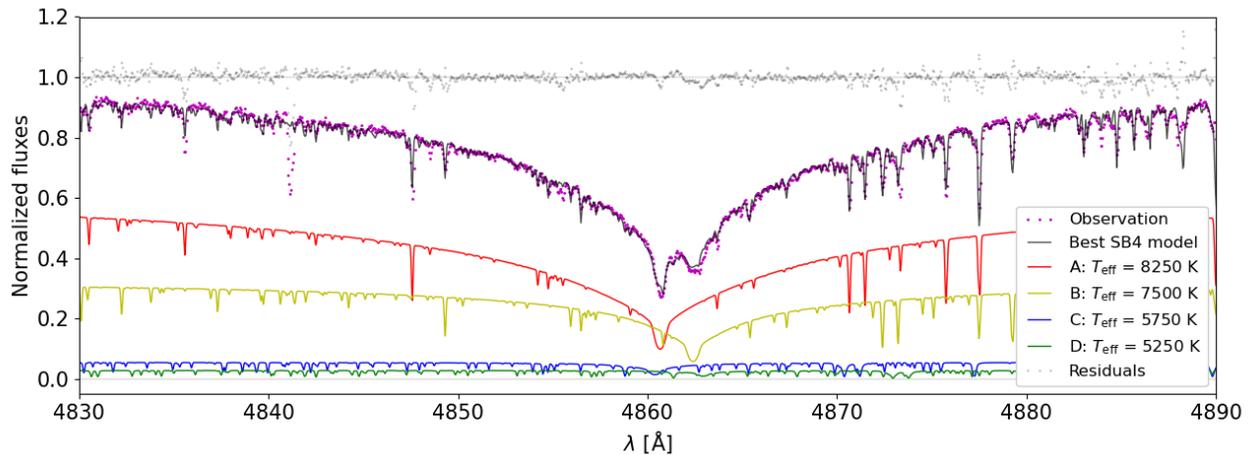

**Figure 1: Spectral fitting of the HRS/SALT spectra (taken on 2018-12-31).** The observed spectrum is shown as magenta dots and the best SB4 template is shown as a black solid line. The four individual component–synthetic spectra are displayed as labelled in the legend. For components C and D, the temperatures adopted in this figure correspond to those of the closest models in the grid of step 250 K. The residuals (observed–synthetic) are shown at the top. The strong line close to the middle is the Balmer line Hβ.

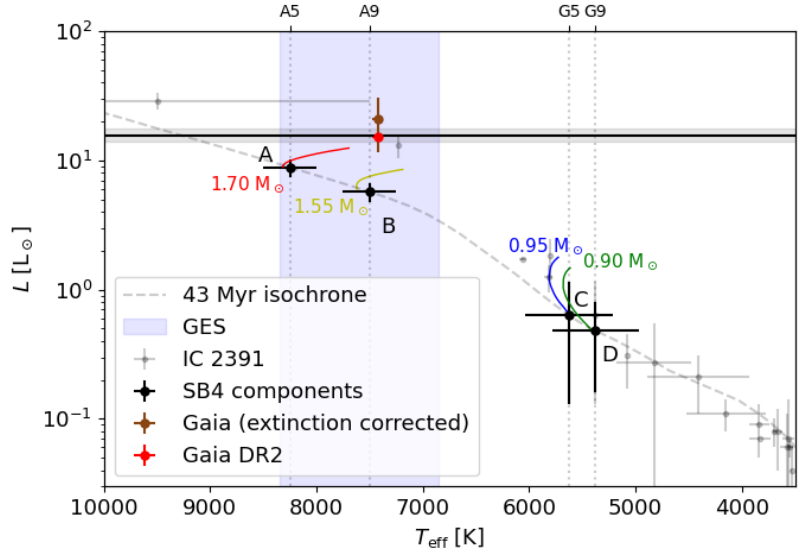

**Figure 2: Location of HD 74438 in the HR diagram.** Black dots mark the positions of the four individual components placed on the main sequence (dashed grey line), with $T_{eff}$ as derived from spectral fitting (vertical dotted lines with associated spectral type on top), to be compared with the temperature of the unresolved system derived by the Gaia-ESO Survey (the blue shaded area representing the 1σ confidence interval). The horizontal black error bars are the $T_{eff}$ uncertainties computed by adding quadratically the error on the mean (since the $T_{eff}$ determination was performed on two spectra) and the grid step. The vertical black error bars are the uncertainties on luminosities deduced from the uncertainties on the component temperatures and cluster age. The combined luminosity derived in this work is plotted as the horizontal black line (the grey shaded area representing the 1σ confidence interval). This combined luminosity can be compared to (i) the luminosity listed in Gaia DR2 ([18], red dot, plotted at the temperature also listed in [18]) and (ii) the one derived from parallaxes, bolometric correction and extinction correction (brown dot). The dashed grey line and colored lines are the theoretical stellar isochrone at 43 My and for illustrative purposes the evolutionary tracks [17] for masses close to the derived ones. Grey dots with error bars represent known members of the IC 2391 open cluster [14].

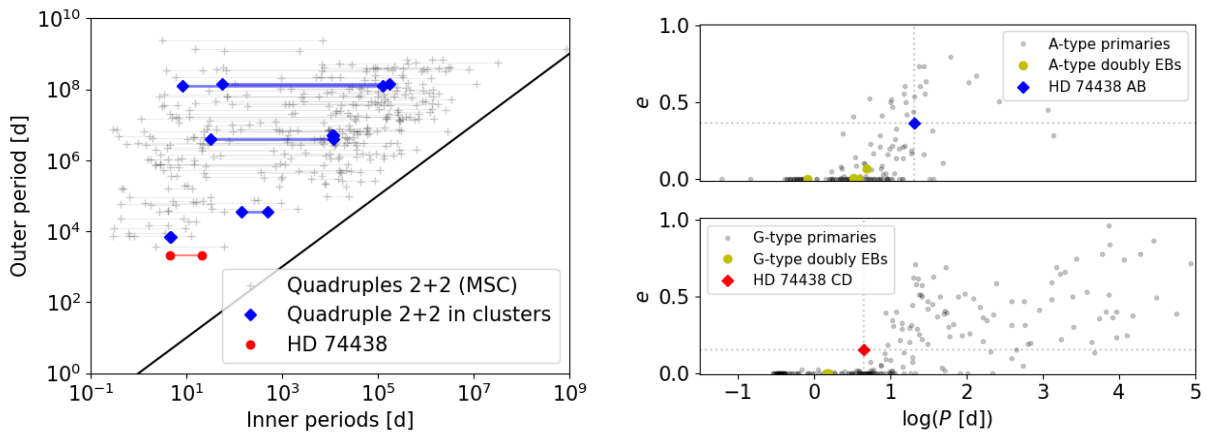

**Figure 3. Comparison of the orbital parameters of HD 74448 with the ones of other quadruple and binary systems.** Left: Comparison with other known 2+2 quadruples in clusters. Grey plusses are the 2+2 quadruples from the Multiple Star Catalogue [22, not volume complete]. We highlighted blue quadruples belonging to clusters as cross-matched with SIMBAD. HD74438 appears to have the smallest

outer period. The solid line represents the condition for 2+2 quadruples to be dynamically stable [3]. Right: Location of the two inner binaries (AB in blue, CD in red) in the eccentricity-period diagram, compared with A-type primaries (matching the AB pair; grey dots in the top panel) and G-type primaries (matching the CD pair; grey dots in the bottom panel) of the SB9 catalogue [32] of spectroscopic binaries and doubly eclipsing binaries [33] in yellow circles.

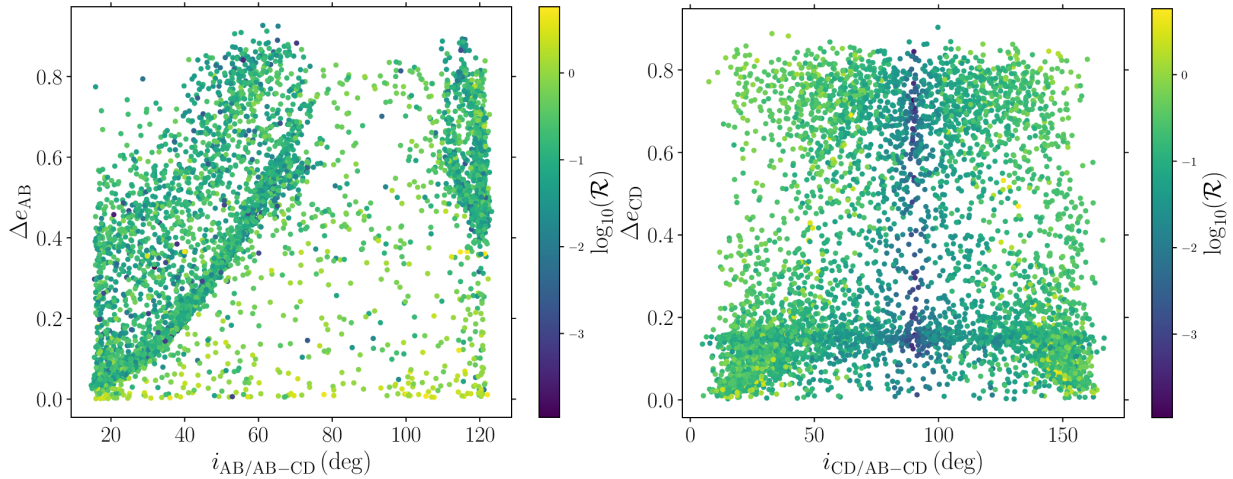

**Figure 4. MSE simulations of the HD 74438 future.** The figures show, for the surviving systems, the amplitude of eccentricity oscillations during the integrated time (i.e., up to 10 Gy) as a function of the initial mutual inclination. Colour-coded are the initial values of $R$ (the ZLK time-scales **ratio** for inner-to-outer orbit pairs; Equation 32 of [29]). In the case of the $e_{AB}$ amplitude (left panel), there is a large gap near 90° inclinations, since these have merged during their previous evolution. The $e_{CD}$-plot (right panel) shows that high eccentricity amplitudes are possible for a large range of initial mutual inclinations $\Phi_{CD/AB-CD}$. Due to quadruple dynamics, the range of mutual inclination angles is much enhanced, as can be seen by noting the colours: $R$'s close to unity are required to get high amplitudes for small mutual inclinations, whereas if $\Phi_{CD/AB-CD}$ is close to 90°, high amplitudes can also be attained regardless of $R$ values. The range of mutual inclinations considered in the simulations represented in the above panels correspond to those predicted by Equation (13) for one of the four possible combinations of inclinations (Supplementary Table 3), namely $i_{AB}$ = 52.5°, $i_{CD}$ = 84.0°, and $i_{AB-CD}$ = 73.2°.

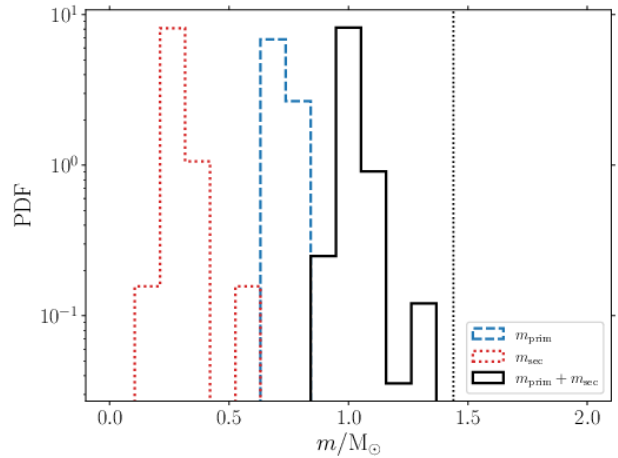

**Figure 5. Distribution of merging WDs in our simulations.** The total mass of the two merging WDs (solid black line), as well as the individual two WD masses resulting, most of the time, from the merger of the AB pair (blue dashed) and CD pair (red dotted lines). The black vertical dotted line indicates the Chandrasekhar mass of 1.44 M$_\odot$.

# Extended Data

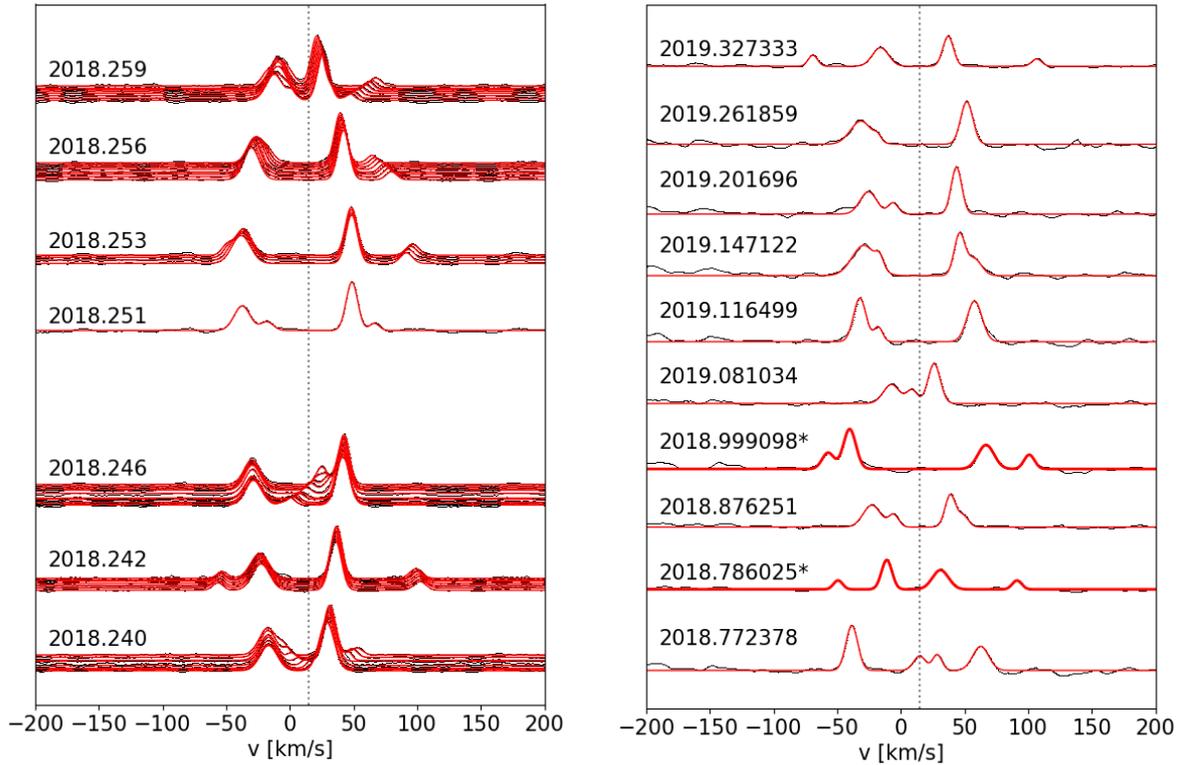

**Extended Data Figure 1. CCFs computed from high-resolution spectra of HD 74438 at different epochs as labeled (black lines) and multi-Gaussian fits (red lines).** The vertical dotted line is the mean velocity of IC 2391. Left: HERCULES/UCMJO CCFs (with multiple spectra observed in some of the nights). The time series covers one week and the vertical position of the spectra scales with time. Right: same as left for HRS/SALT CCFs. The time series covers 7 months. The vertical shift of the spectra is arbitrary. Spectral fitting was performed using the two spectra marked with an asterisk and plotted in bold.

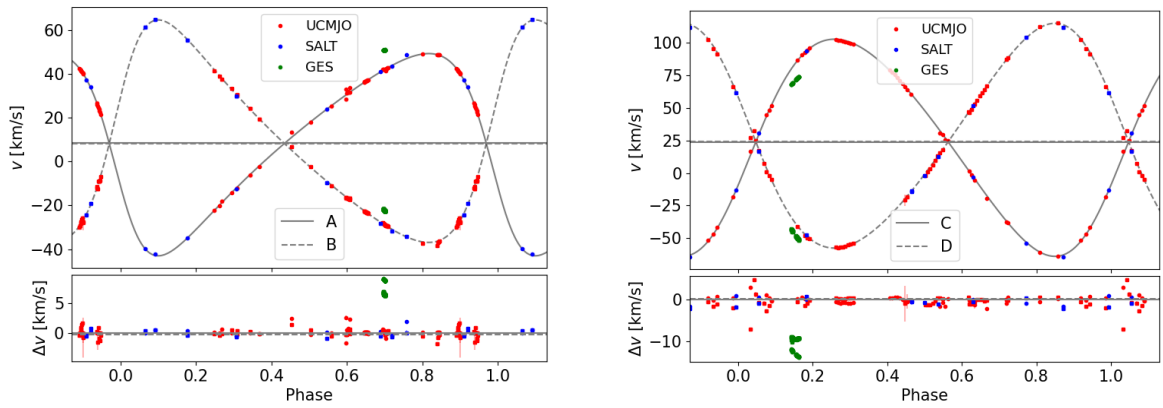

**Extended data Figure 2. RV solutions using HERCULES/UCMJO and HRS/SALT data points for the brightest AB pair (left) and for the faintest CD pair (right).** The GES data points are not included in the computation of the orbital solutions because taken 5 y before, showing the gravitational effect between the two inner pairs.

|  | **A-B** | **C-D** | **AB-CD** |
|---|---|---|---|
| *P* [d] | 20.5729 ± 0.0003 | 4.4243 ± 0.0001 | 2074.2 ± 3.5 |
| *e* | 0.3692 ± 0.0001 | 0.1535 ± 0.0003 | 0.458 ± 0.015 |
| $\omega_1$ [rad] [a] | 1.8780 ± 0.0003 | -1.946 ± 0.002 | 0.185 ± 0.039 |
| $T_0$ - 2400000 [d] | 58 605.9 ± 0.1 | 58 684.5 ± 0.1 | 59165.8 ± 5.1 |
| $v_0$ [km s$^{-1}$] | - | - | 14.5 ± 0.2 |
| $K_1$ [km s$^{-1}$] [b] | 45.81 ± 0.09 | 83.2 ± 0.1 | 12.8 ± 0.3 |
| $K_2$ [km s$^{-1}$] [b] | 50.77 ± 0.09 | 85.5 ± 0.1 | 18.5 ± 0.4 |
| $q_{dyn}$ | 0.902 ± 0.002 | 0.973 ± 0.002 | 0.692 ± 0.003 |
| $q_{spec}$ | 0.91 ± 0.05 | 0.91 ± 0.22 | 0.58 ± 0.11 |
| *i* [°] | (52.5 or 127.5) ± 1.5 | (84.0 or 96.0) ± 0.9 | (73.2 or 106.8) ± 2.7 [c] |
| *a* [au] | 0.215 ± 0.002 | 0.0681 ± 0.001 | 5.54 ± 0.04 |
| Ω [°] | - | - | 333° or 274° |
| $\mu"_{phot}$ [mas/y] | - | - | 13.0 |

(a) The argument of periastron ω corresponds to the spectroscopic orbit of the brightest component around the centre of mass.
(b) $K_1$ refers to $K_A$, $K_C$ or $K_{AB}$ while $K_2$ refers to $K_B$, $K_D$ or $K_{CD}$
(c) The solution with $i_{AB-CD}$ = 73.2° is favoured by the Gaia proper motion data (see text).

**Extended Data Figure 3. Orbital parameters of the two inner orbits (A-B and C-D) and the outer one (AB-CD) of the quadruple 2+2 system HD 74438.** *P* is the orbital period, *e* the eccentricity, ω the periastron argument, $T_0$ the julian date at periastron, $v_0$ the center-of-mass velocity, $K_1$ and $K_2$ the radial velocity amplitudes of the primary and secondary in each orbit. $q_{spec}$ and $q_{dyn}$ are the spectroscopic and dynamical mass ratios. *i* is the inclination of the orbit on the sky, *a* the semi-major axis of the orbit around the center of mass, $\mu"_{phot}$ the proper motion of the photocentre and Ω the argument of the ascending node.

| Component | A | B | C | D | Unresolved |
|---|---|---|---|---|---|
| Spectral Type | A5 | A9 | G5 | G9 | - |
| $T_{eff}$ [K] | 8250 ± 250 | 7500 ± 250 | 5625 ± 410 | 5375 ± 410 | - |
| $L$ [$L_\odot$] | 8.87 ± 1.40 | 5.72 ± 0.95 | 0.64 ± 0.51 | 0.48 ± 0.32 | 15.71 ± 1.80 |
| $R$ [$R_\odot$] | 1.46 ± 0.15 | 1.42 ± 0.15 | 0.84 ± 0.36 | 0.80 ± 0.29 | - |
| $M$ [$M_\odot$] (a) | 1.70 ± 0.06 | 1.54 ± 0.06 | 0.96 ± 0.14 | 0.87 ± 0.14 | 5.07 ± 0.22 |
| $M$ [$M_\odot$] (b) | 1.64 ± 0.06 | 1.48 ± 0.06 | 1.09 ± 0.04 | 1.06 ± 0.04 | 5.27 ± 0.10 |

(a) Masses derived from the location of the stars in the HR diagram.
(b) Masses derived from Equations (5) and (6), adopting the value of $M_A$ from (a).

**Extended Data Figure 4. Astrophysical parameters of the quadruple system HD 74438 and its components.** $T_{eff}$ is the effective temperature, $M$, $L$ and $R$ are the mass, the luminosity and the radius in solar units.

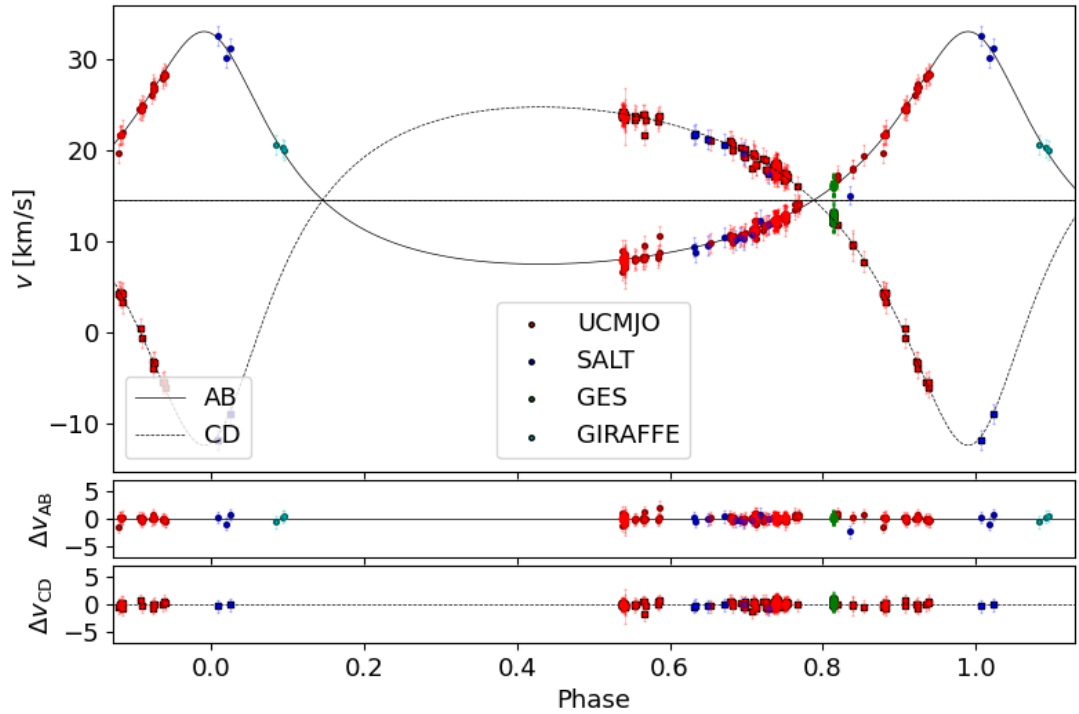

**Extended Data Figure 5. RV solutions of the wide pair AB-CD using center of mass RVs of the AB and CD pairs.**

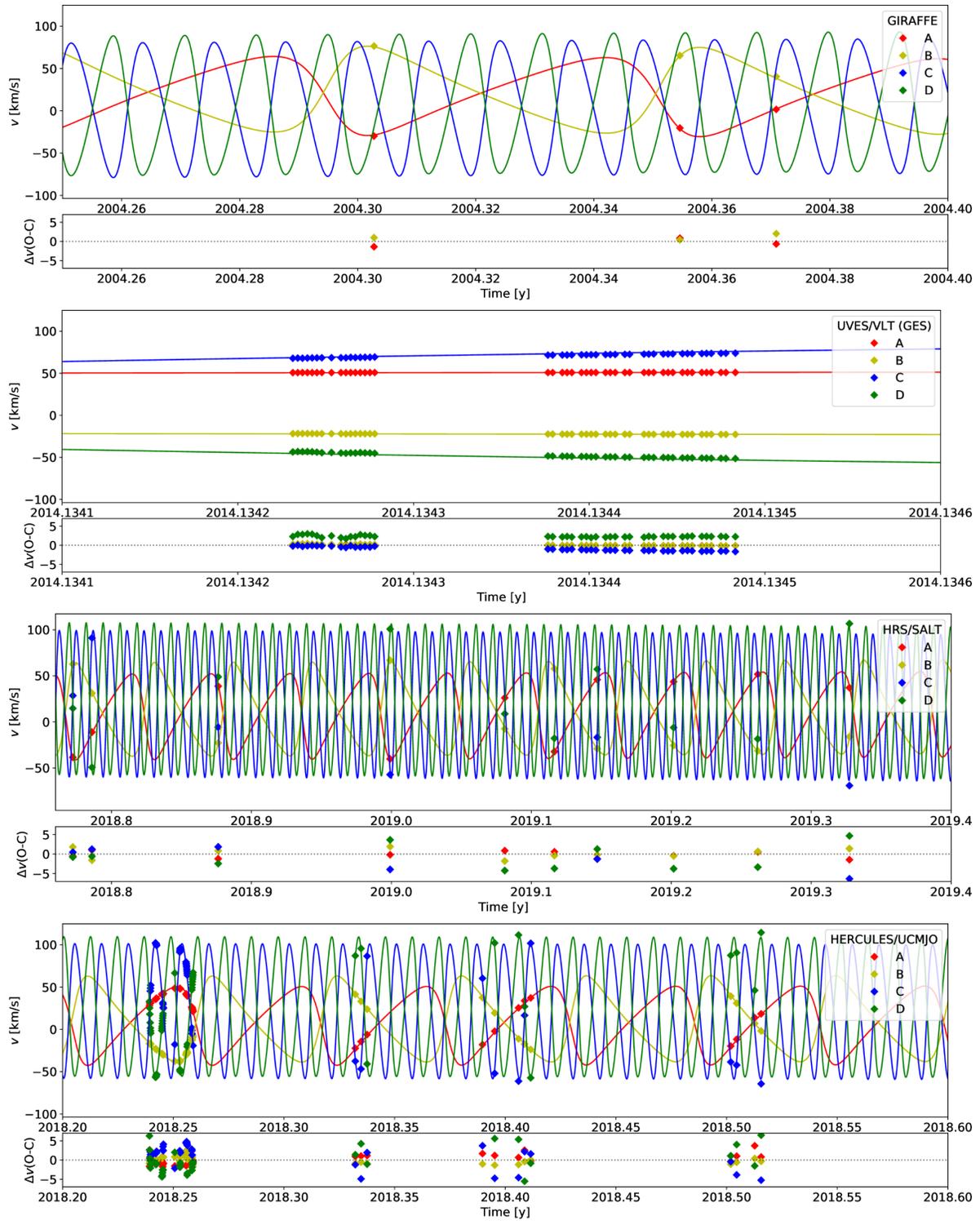

**Extended Data Figure 6. Orbital solutions from Extended Data Figure 3 and measured RVs for the four components of the SB4**. Top: GIRAFFE data, middle top: UVES/VLT from GES data, middle bottom: HRS/SALT data, bottom: HERCULES/UCMJO data. Not all the RVs are presented here.

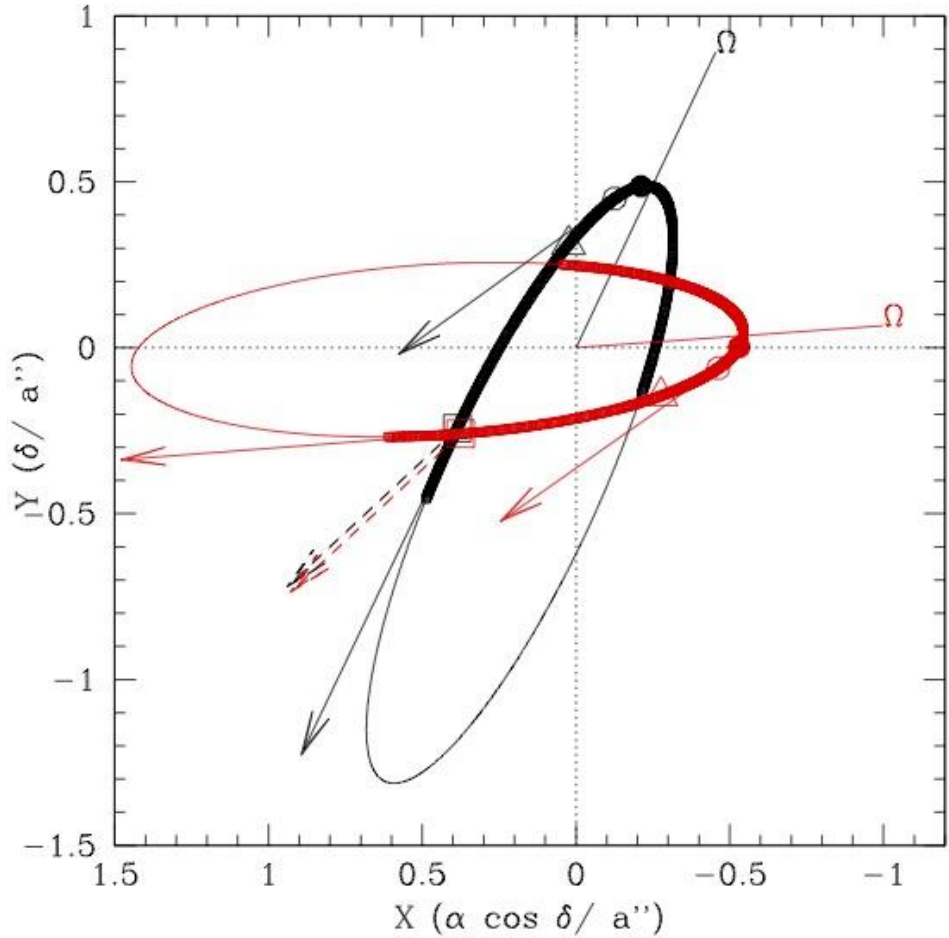

**Extended Data Figure 7. The two possible orbits of the photocentre of the AB pair around the centre of mass of the AB - CD system (in units of the semi-major axis $a''_{AB-CB}{}^{phot}$).** The black solid line corresponds to the orbit of inclination 73.2°, whereas the red solid line corresponds to the orbit with inclination 106.8°. On these orbits, following the direction of orbital motion, the solid circle marks the periastron, the open circle marks the epoch 2015.3 (April 24) when the predicted orbital velocity projected on the plane of the sky (solid-line arrow) matches the Gaia DR2 proper motion (differential with respect to the cluster, and expressed in units of the proper-motion modulus $\mu_{AB-CD}{}^{phot}$) averaged over the Gaia DR2 time span (represented by the thick part of the orbit), the triangle marks the average Gaia DR2 epoch (2015.5), the open square marks the Gaia eDR3 epoch (2016.0). For this epoch, the solid-line arrow marks the predicted orbital motion projected on the plane of the sky, whereas the dashed arrow corresponds to the Gaia eDR3 proper motion (differential with respect to the cluster). The counter-clockwise evolution of the observed proper motion (Gaia DR2 and Gaia eDR3) favours the orbit with 73.2° inclination (black line).

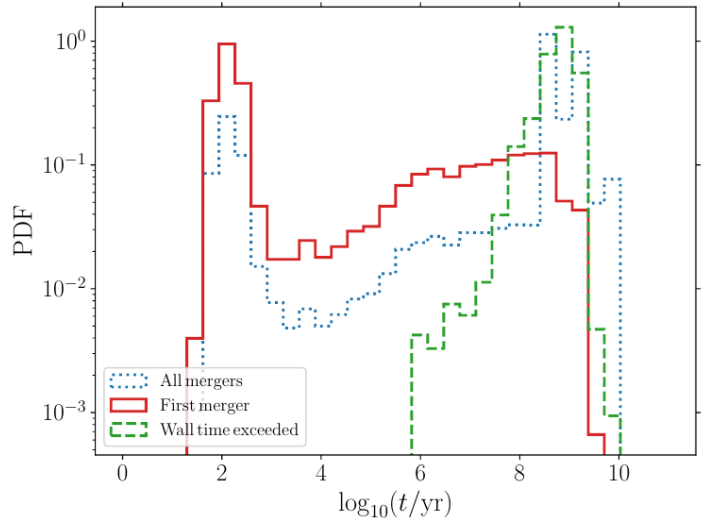

**Extended Data Figure 8. Probability distributions of merger times in the simulations.** The distribution of times of all merger events is displayed with the blue dotted lines, and the times of the first merger in the system (if applicable) with the red solid line. The distributions of the maximum age reached by the system for the systems in which the maximum wall time was exceeded is shown with the green dashed line.

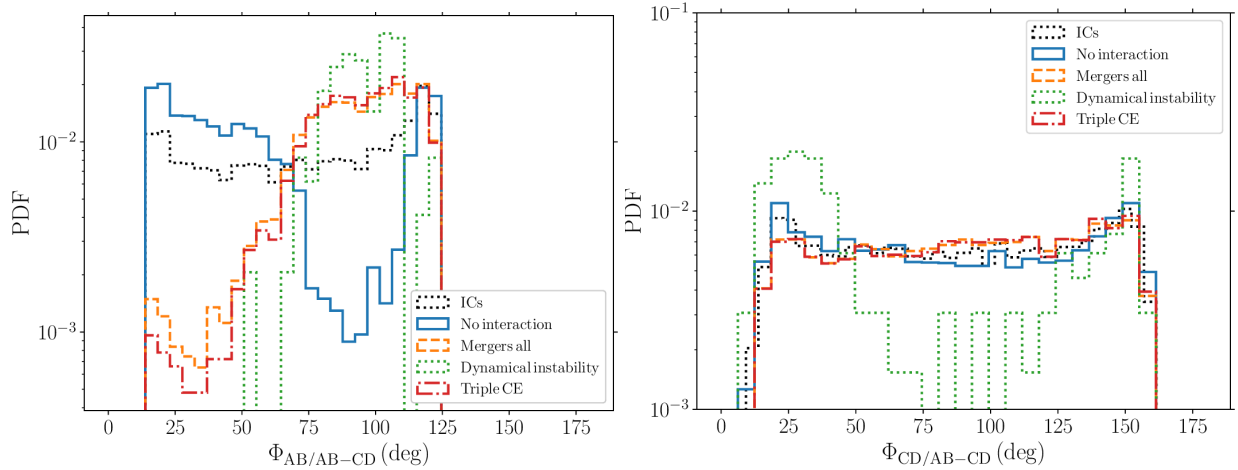

**Extended Data Figure 9**. **Probability distributions of the current mutual inclinations (left panel: $\Phi_{AB/AB-CD}$ ; right panel: $\Phi_{CD/AB-CD}$) leading to different events in our Monte Carlo simulations (as described in the legends).** The curve labelled ICs corresponds to the distribution of the initial mutual inclinations (encompassing all possible outcomes). This distribution is obtained from the observed values of the individual orbital inclinations on the sky, complemented by flat distributions for the longitudes of the unknown ascending nodes (Equation 12).

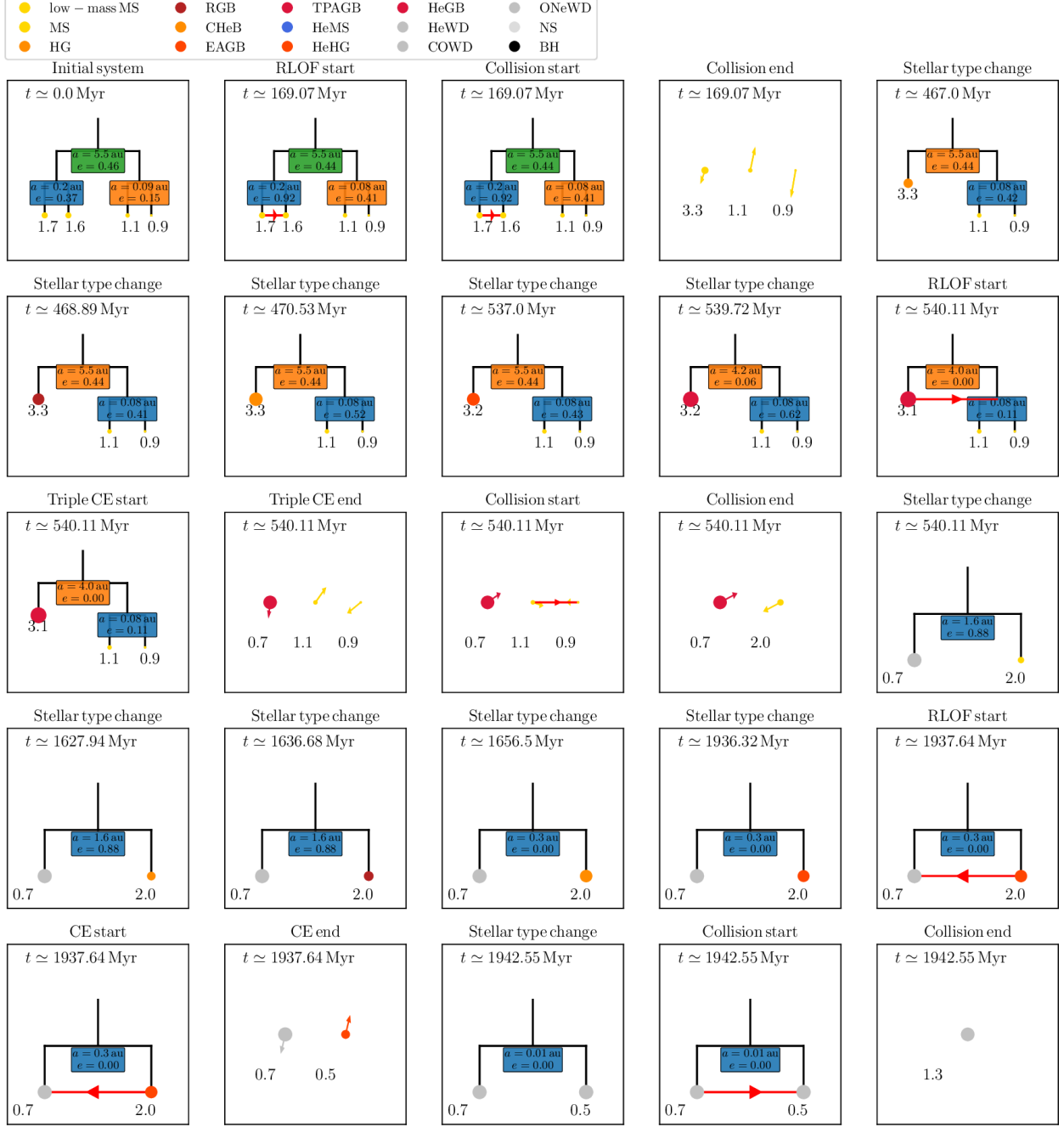

**Extended Data Figure 10. One possible future evolution of HD 74438.** Each panel shows an event of interest as labelled at the top, with the time indicated in each subpanel and the system represented schematically in a so-called mobile diagram [35], with orbital parameters (semimajor axes *a* and eccentricities *e*) and masses (in units of $M_\odot$) indicated. The meaning of the dot colours (representing the stars) is indicated in the legend at the top. The quadruple experiences unstable RLOF, triple CE and merger events, leaving a WD remnant with a sub-Chandrasekhar mass of 1.3 $M_\odot$.

# Supplementary information

## Notes

The statistics of higher-order stellar systems are still very uncertain [1]. Recent large spectroscopic surveys have harvested many spectroscopic multiple systems with two and three components [2, 3] but very few [4] with four components (SB4).

**Already known SB4.** Quadruple systems with main-sequence components represent about 4% of all stellar systems in the Solar Neighbourhood. About 425 quadruples are known [5] and the 2+2 hierarchy appears to be more observed than 3+1 and trapezium systems. Nevertheless quadruple *spectroscopic* binaries are rather rare. To our knowledge, only nine SB4 were previously found, none of them in a cluster, all of them dimmer than HD 74438, and all of them showing eclipses. Four of them (6., 7., 8. and 9.) have reached the degree of characterization of HD74438, but apparently without ZLK cycles at work. We report them below by chronological order of characterisation:

1. BD−22°5866: K and M binaries with V = 10.4. The K binary is also an EB with a 2.2 d period. No spectral decomposition [6]; unknown inclinations.
2. V994 Her: SB4 composed of 2 pairs of eclipsing binaries: (B8V+A0V) and (A2V+A4V) with 2.1 and 1.4 d periods [7]. The outer period is 2.9 y [8].
3. KIC 4247791: SB4 system with two eclipsing binaries made of 4 F-type stars (F0, F2, F7, F8) with 4 d periods for each EB and V = 11.6. No spectral decomposition [9];
4. KIC 7177553: SB4 system consisting of two eccentric binaries with similar periods of about 17 d where one of the two binaries is eclipsing, V = 11.3. The four components are G-type stars of similar masses [10].
5. EPIC 220204960: SB4 system with two interacting eclipsing binaries made of 4 M stars with periods of 13-14 d with an outer period of about 1 y, and V = 12.7 [11].
6. V482 Per: SB4 system (B9, A1, A7, A7) with 2 EB with 2.4 and 6 d with an outer period of 16.6 y, and V = 10.3 [12].
7. VW LMi: SB4 system (F-G spectral types) which is the tightest quadruple system with 2+2 hierarchy yet discovered, with 0.48, 7.93 and 355 d periods and V = 8.0 [13].
8. CzeV1731: SB4 system with 2 twin eclipsing binaries and an outer period estimated at 34 y [14].
9. TIC 454140642: a coplanar SB4 system with 2 eclipsing binaries with V = 10.4. The outer period is estimated at 432 d [15].

More SB4 candidates start to emerge from large-scale spectroscopic surveys like APOGEE [16] and GALAH [17] but need follow-up observations to confirm their quadruple nature and, ultimately, characterize their orbital and astrophysical parameters.

**Other 2+2 quadruple (but not SB4) in open clusters.** Our literature search (from the Multiple Star Catalogue [5] and the doubly eclipsing systems [18]) has revealed the following quadruple systems within an open cluster (OC):

- μ Ori, a member of the Hyades, is one of the brightest quadruples, with both orbits and inclinations known, as well as masses and luminosities derived from interferometry [19];
- HD 5980 (WN4+O7 I:), a member of the young OC NGC 346 in the SMC, which has been the topic of a detailed spectral analysis, but its quadruple nature is uncertain [20];
- The four other quadruples in the Hyades, as well as HD 46180 in NGC 224, and HY and KT Vel in IC 2391, have not reached the same level of characterisation as μ Ori;
- OGLE LMC-ECL-02903 is a doubly eclipsing binary in the Large Magellanic Cloud open cluster KMH 283; the components are only characterized from the eclipse model fit using the photometric data of the OGLE survey [18].

**Higher-order multiples in open clusters.** Higher order multiples are difficult to discover and the ones reported in the MSC [5] mainly result from incidental findings. To date, 70 quintuples and 18 sextuples have been reported, with few of them located in OC and molecular clouds. Five septuples have also been reported with most of them being part of moving groups or associations. Due to the strong statistical biases, any conclusion on the frequency of multiples in the field and in OC is still premature. Large on-going and future surveys will allow us to increase the statistics of these interesting high-order multiples.

| Instrument | Resolving power ($\lambda/\Delta\lambda$) | Spectral coverage (Å) | Number of exposures | Epoch range |
|---|---|---|---|---|
| GIRAFFE/VLT | 22 000 | [3 850, 4 050] | 3 | 25 d |
| UVES/VLT | 47 000 | [4 800, 6 200] | 45 | 2.5 h |
| HERCULES/UCMJO | 41 000 | [4 160, 7 635] | 199 | ~2.3 y |
| HRS/SALT | 65 000 | [3 830, 8 775] | 14 | ~2 y |

**Supplementary Table 1. Summary of the spectroscopic observations.**

| Catalogue | HD 74438 (Tycho-1) | HD 74438 (Tycho-2) | HD 74438 (Gaia DR2) | HD 74438 (Gaia eDR3) | Cluster center |
|---|---|---|---|---|---|
| Mid-epoch (y) | 1991.25 | 1991.23 | 2015.5 | 2016.0 | - |
| Time span (y) | 3.5 | ~ 100 | 1.8 | 2.8 | - |
| $\mu_\alpha^*$ (mas y$^{-1}$) | -33.5 ± 4.0 | -23.4 ± 1.5 | -15.895 ± 0.122 | -17.74 ± 0.09 | -24.93 ± 0.08[13.] |
| $\mu_\delta$ (mas y$^{-1}$) | +27.8 ± 3.4 | +23.7 ± 1.4 | +17.256 ± 0.131 | +17.06 ± 0.10 | +23.26 ± 0.11[13.] |
| $\Delta\mu_\alpha^*$ (mas y$^{-1}$) | -8.57 ± 4.0 | 1.53 ± 1.5 | 9.03 ± 0.14 | 7.19 ± 0.12 | - |
| $\Delta\mu_\delta$ (mas y$^{-1}$) | 4.54 ± 3.4 | 0.44 ± 1.4 | -6.00 ± 0.17 | -6.20 ± 0.15 | - |
| $\dot{X} = \Delta\mu_\alpha^* / \mu''_{AB-CD,phot}$ | -0.659 | 0.118 | 0.694 | 0.553 | - |
| $\dot{Y} = \Delta\mu_\delta / \mu''_{AB-CD,phot}$ | 0.349 | 0.034 | -0.461 | -0.477 | - |
| $\dot{X}^2 + \dot{Y}^2$ | 0.556 | 0.015 | 0.694 | 0.533 | - |

**Supplementary Table 2. Proper motions in right ascension (RA, including the cos *δ* factor) and declination (*δ*).** The differential proper motion $\Delta\mu_\alpha^*$ and $\Delta\mu_\delta$ denote the differential stellar motion with respect to the cluster. The differential proper motions from Tycho-1, Gaia DR2, and Gaia eDR3 reflect the orbital motion of the ABCD photocentre around the centre of mass of the cluster.

| $I$ [°] | | | $\Phi_{AB, AB-CD}$ or $\Phi_{CD, AB-CD}$ [°] | |
|---|---|---|---|---|
| AB | CD | AB-CD | min | max |
| 52.5 | - | 73.2 | 20.7 | 125.7 |
| 52.5 | - | 106.8 | 54.3 | 159.3 |
| 127.5 | - | 73.2 | 54.3 | 159.3 |
| 127.5 | - | 106.8 | 20.7 | 125.7 |
| - | 84.0 | 73.2 | 10.8 | 157.2 |
| - | 84.0 | 106.8 | 22.8 | 169.2 |
| - | 96.0 | 73.2 | 22.8 | 169.2 |
| - | 96.0 | 106.8 | 10.8 | 157.2 |

**Supplementary Table 3. The limits on the mutual inclinations $\Phi$ as derived from Equation (13) for the different possible combinations of the orbital inclinations.**

| Description | Fraction | Final outcomes | Fraction |
|---|---|---|---|
| No interaction | 0.535 ± 0.007 | Quadruple | 0.535 ± 0.007 |
| Mergers all | 0.465 ± 0.007 | Single | 0.236 ± 0.005 |
| CE | 0.298 ± 0.005 | Triple | 0.117 ± 0.003 |
| Collision | 0.461 ± 0.007 | Two Single | 0.058 ± 0.002 |
| Dynamical instability | 0.010 ± 0.001 | Binary | 0.030 ± 0.002 |
| Triple RLOF | 0.363 ± 0.006 | Binary+Single | 0.024 ± 0.002 |
| Triple CE | 0.360 ± 0.006 | | |
| **Number of final remnants** | **Fraction** | **Triple CE outcomes** | **Fraction** |
| 1 | 0.236 ± 0.005 | Triple | 0.036 ± 0.003 |
| 2 | 0.088 ± 0.003 | Merger(s) | 0.705 ± 0.014 |
| 3 | 0.141 ± 0.004 | Binary+Single | 0.152 ± 0.006 |
| 4 | 0.535 ± 0.007 | Indeterminate | 0.107 ± 0.005 |

**Supplementary Table 4. Outcome fraction per physical process (several processes can actually occur for a given simulated system), number of remnants and final outcomes of MSE simulations.** Statistical error bars are given (based on Poisson statistics). 'Mergers all' means that any kind of merger occurred at any point, irrespective of whether or not an ejection event happened. Dynamical instability represents instability triggered by stellar evolution (e.g. orbital expansion due to wind mass loss). In some cases, the outcome was indeterminate from the simulation data in relation to the CPU wall time being exceeded.